\documentclass[12pt]{article}
\def\beqar {\begin{eqnarray}}
\def\eeqar {\end{eqnarray}}
\def\beq {\begin{equation}}
\def\eeq {\end{equation}}
\def\ra {{\rangle}}
\def\la {{\langle}}
\def\half {{\textstyle{1\over 2}}}
\def\Tr {{\rm Tr}}
\def\tr {{\rm tr}}

\def\del {{\partial}}

\def\ep {{\epsilon}}

\def \D {{\cal D}}

\def\no2 {{\textstyle{n\over 2}}}

\begin{document}

\begin{titlepage}
\null\vspace{-62pt}

\pagestyle{empty}
\begin{center}
\rightline{CCNY-HEP-03/5}

\vspace{1.0truein}
{\large\bf Quantum Hall effect on $S^3$, edge states
and fuzzy $S^3/{\bf Z}_2$}\\
\vskip .3in
\vspace{.5in}V.P. NAIR $^{a b }$
{\footnote{vpn@sci.ccny.cuny.edu}}
  and S. RANDJBAR-DAEMI $^{b }$
{\footnote
{daemi@ictp.trieste.it}}\\
\vspace{.3in}{\it $^a$ Physics Department\\
City College of the CUNY\\
New York, NY 10031}\\
\vspace{.1in}{\it $^b$Abdus Salam International Centre for
Theoretical
Physics\\
Trieste, Italy}\\
\vspace{0.1in}
\end{center}
\vspace{0.5in}

\centerline{\bf Abstract}

We analyze the Landau problem and quantum Hall effect
on $S^3$ taking a constant background field proportional
to
the spin connection on $S^3$. The effective strength
of the field can be tuned by changing
the dimension of the representation
to which the fermions belong.
The effective action for the edge
excitations of a quantum Hall droplet in the limit of
a large number of
fermions is obtained.
We find that the appropriate space for many of
these considerations is
$S^2 \times S^2$, which plays a role
similar to that of ${\bf CP}^3$
{\it vis-a-vis} $S^4$.
We also give a method of representing the
algebra of functions on fuzzy $S^3/{\bf Z}_2$
in terms of finite dimensional matrices.

\baselineskip=18pt

\end{titlepage}

\hoffset=0in
\newpage
\pagestyle{plain}
\setcounter{page}{2}
\newpage
\section{Introduction}
Quantum Hall effect (QHE)
has been analyzed on many higher dimensional spaces,
motivated by the original
analysis by Zhang and Hu
\cite{HZ}, who considered the Landau problem for charged
fermions on $S^4$ with a background magnetic field which
is the standard $SU(2)$ instanton.
In the classic two-dimensional QHE, one can easily see
that
a droplet of fermions occupying a certain area behaves as
an incompressible fluid. The low energy excitations
of such a droplet are then
area-preserving deformations which behave as massless
chiral
bosons. This suggested that for a droplet
in QHE on $S^4$
the edge excitations could lead to higher spin massless
fields, in particular the graviton.
This might then provide an interesting new way
to a
quantum description of a graviton in four dimensions.
Although
this has not been born out, edge excitations for
a quantum Hall droplet lead to an interesting class of
field theories which are intimately linked to
the geometry of the underlying space
and so they should merit further study.

There have been a
number of papers extending the original
idea of Zhang and Hu to other even dimensional spaces
\cite{KN1}- \cite{pol}.
The effect has been analyzed and edge excitations
obtained
on even dimensional complex projective
spaces
${\bf CP}^k$ which allow both abelian and
nonabelian background fields \cite{KN1,KN2}.
The model on $S^4$ can be understood as QHE
with a $U(1)$ background magnetic field
in ${\bf CP}^3$, because the latter is an
$S^2$-bundle over $S^4$.
This point of view has been
investigated in \cite{KN1, berneveg, zhang, ners}
and the effective action for the edge states obtained in
\cite{KN2}.

Edge excitations for droplets in $R^4$ with
$U(1)$ and $SU(2)$ backgrounds,
which would correspond to a flat
space limit of
the Zhang-Hu model,
were studied in \cite{pol}.
Since a droplet of finite volume
is topologically
a neighbourhood in $R^4$,  the analysis in \cite{pol}
leads to many
of the generic features
of the edge excitations. For example,
the effective theory of edge excitations
is essentially an
infinite collection
of one-dimensional theories and also it does not
have full Lorentz
invariance.

All the analyses mentioned have been on even dimensional
spaces. For any coset of the $G/H$ type, where
$G$ is a Lie group and $H$ a compact subgroup
(of dimension $\geq 1$),
there is always the analogue of a constant background
field;
it is given by the spin connection on $G/H$.
Thus it is possible to consider QHE on such spaces
taking the gauge field to be proportional to the
spin connection. For two dimensions,
one can consider $SU(2)/U(1)$ which admits a constant
$U(1)$ background field and leads to the usual QHE.
In three dimensions, the simplest case to
consider is $S^3 = SU(2)\times SU(2)/SU(2)$;
one can get an $SU(2)$ background field.
When we go to four dimensions, for $S^4$, the isotropy
group is $H= SO(4)\sim SO(3) \times SO(3)$ giving the
possibility
of selfdual and antiselfdual fields, the instantons.
For ${\bf CP}^2 = SU(3)/U(2)$, one can get either
abelian or $SU(2)$ background fields.
Considering models of
increasing complexity,
we see that an interesting and simple case, namely
$S^3$ has not yet been analyzed and this is the subject
of the present paper.
We will construct Landau levels, analyze droplets for
the lowest Landau level and obtain the effective action
for the edge states. There is also an interesting
connection
between the Landau problem and fuzzy spaces, the algebra
of
functions on the latter being realized in terms of
operators on
the lowest Landau level. This is not quite true
for $S^3$, but our analysis does lead to
an interesting definition of fuzzy $S^3 /{\bf Z}_2$.

The paper is organized as follows. In section 2 we
give the Landau levels. The action for edge excitations
is derived in section 3. Section 4 deals with fuzzy
$S^3/{\bf Z}_2$.
We conclude with a short discussion.

\section{Landau levels for $S^3$}

As in the previous analyses, the calculation is
facilitated if the
space of interest is a coset of groups, so we
begin with the observation that
the three-sphere
$S^3$ may be considered as $ G/H = SU(2) \times SU(2)/
SU(2)_{diag}$, where $SU(2)_{diag}$ is the diagonal
subgroup of the two $SU(2)$ groups.
Functions $f(u, u' )$ on $SU(2) \times SU(2)$ can be
expanded as
\beq
f(u, u')= \sum f_{mnm'n'}^{(l,l')}~{\cal D}_{mn}^{(l)} (u
)
~{\cal D}_{m'n'}^{(l')}(u' )
\label{1}
\eeq
where ${\cal D}_{mn}^{(j)} (g)$ are the Wigner
${\cal D}$-functions on
$SU(2)$, defined by
\beq
{\cal D}_{mn}^{(j)}= \la jm\vert {\hat u} \vert jn\ra
\label{2}
\eeq
where ${\hat u} = \exp (i J\cdot \theta )$ is the operator
corresponding to the group element $\exp (i t_a
\theta^a)$,
with $t_a = \half \sigma_a$, $a=1,2,3$ and $\sigma_a$ are
the Pauli
matrices. We use $u$ to denote elements of the first
$SU(2)$ and
$u'$ to denote elements of the second $SU(2)$.
From (\ref{2}), functions, vectors, tensors, etc. on $S^3$ may be
constructed by suitably restricting the choice
of representation
of $SU(2)_{diag}$. Let $R_a$ and $R'_a$
denote generators
of the first and second $SU(2)$'s in
$G$ respectively which correspond to right
translations of $u$ and $u'$ respectively.
\beqar
R_a ~u = u ~t_a, &&  R'_a ~u' = u'~ t_a\nonumber\\
R_a {\cal D}(u)_{mn} = {\cal D}(u t_a)_{mn},&&
R'_a {\cal D}(u')_{mn} = {\cal D}(u' t_a)_{mn}
\label{3}
\eeqar
Then $J_a = R_a +R'_a$ are
the generators of the diagonal subgroup
$SU(2)_{diag}$.
Scalar uncharged functions on $S^3 =SU(2)\times SU(2)'/
SU(2)_{diag}$ must be invariant under $J_a$, by
definition, since
we are dividing out by $SU(2)_{diag}$, and thus we can
obtain them from
(\ref{1}) by choosing combinations on which $J_a =0$.
Derivatives on
functions can be represented as
\beqar
P_i&=& -i \nabla_i ~\equiv~
{1\over 2r} K_i \nonumber\\
K_i &=& R_i - R'_i
\label{3a}
\eeqar
where $r$ is the radius of the sphere.
The operators $K_i$ obey the commutation rules
\beq
[K_i ,K_j]=  i~\ep_{ijk} ~J_k
\label{4}
\eeq
We are also interested in spinors and vectors and fields
which have
a gauge charge as well; on these the commutator of
covariant derivatives
must go like
$[\nabla_i,
\nabla_j ]\sim R_{ij}^\alpha S_\alpha +F_{ij}^a T_a$,
where
$R_{ij}^\alpha$ is the Riemann tensor on the space of
interest,
$S_\alpha$ is the spin operator on the fields on which
$\nabla_i$ act,
$F_{ij}^a$ is the gauge field strength and $T_a$ are the
gauge group
generators appropriate to the fields on which they act.
For $S^3$ we
have constant Riemannian curvature. As for the background
gauge field,
the natural choice for a constant
background field is to take the gauge potential to be
the spin connection. This field corresponds to fields
in an $SU(2)$ subgroup of the gauge group, given by
$F_{ij}^a = -\ep_{ija}$. This is a fixed field, like the
instanton field analyzed in \cite{HZ}, but one can tune
the coupling to matter fields by assigning
different
representations of the gauge algebra to them.
As for the spectrum and the degeneracy, only the
combination
of field strength and coupling matters.
Our choice of background as well as the effect of Riemann
curvature can be
encoded in the commutation rules
\beq
[K_i ,K_j] = i\ep_{ijk} (S_k +T_k)
\label{5}
\eeq
or in other words, on the matter fields of interest to us,
$J_k =S_k
+T_k$. Now $T_a$  form
an $SU(2)$ subalgebra in the Lie algebra of the gauge
group;
we will refer to this as the isospin group.

For the Landau problem and the construction of the
wavefunctions, we are interested in fermions coupled to
our choice of background. The fields of interest
are thus scalars for the nonrelativistic problem
(neglecting spin) and Dirac fields for the relativistic
case.

Consider first the nonrelativistic case. The Hamiltonian
is
given by
\beq
H= {P^2\over 2M}= {K^2 \over 8Mr^2}
\label{6}
\eeq
$K^2$ is easily calculated from its definition as
\beqar
K^2 &=& (R - R' )^2 = 2 (R^2 + R'^2 ) - (R
+R' )^2 \nonumber\\
&=&2 (R^2 + R'^2)  -J^2\nonumber\\
&=& 2~l(l+1) +2~l'(l'+1) - J(J+1)
\label{8}
\eeqar
The modes can be constructed in terms of the spin-$l$ and
spin-$l'$
representations of $R_i$ and $R'_i$. The lowest combined
angular
momentum must be $J$. This can be achieved in $(2 J+1)$
ways, so
there are $(2 J+1)$ towers of states for each $J$. These
possibilities
can be written as
\beq
l= {q+\mu\over 2},\hskip 1in l'={q-\mu\over 2} +J
\label{9}
\eeq
$\mu =0,1,2,...,2J$ give the $(2 J+1)$ distinct towers.
$q= 0,1,2,\cdots$
give the states in each tower.
In terms of $(q, \mu )$, the energy eigenvalues are given by
\beq
8mr^2 E= K^2 = (q+J+1)^2 +(J-\mu )^2 -1 - J(J+1)
\label{10}
\eeq
In the present case of no spin for the particles,
$J = T$. The levels are labelled by $q$ and the tower
index
$\mu$.
The number of states or degeneracy $d (q, \mu ,J)$
for a
given
$(q, \mu ,J )$ is given by
\beq
d (q, \mu ,J) =(2 l+1)~(2 l'+1)= (q+\mu +1)~(q-\mu +2 J
+1)
\label{10a}
\eeq

For the lowest level, clearly we must have $q=0$.
Further, among the various values of $\mu$,
we must minimize the term $(J-\mu )^2$.
For integer values of $J=T$, this is obtained for
$\mu =J$, which gives
$8mr^2 E = T$,
$d(0, T, T)= (T+1)^2$.
For values of $T$ which are half an odd integer,
two towers with
$\mu = J \pm \half$ are degenerate with
$8mr^2E = T+{1\over 4}$, $d (0, T\pm \half , T)=
(T+{3\over 2})
(T+\half )$.

The wavefunctions for the Landau levels can be constructed
from (\ref{1}). They are given by
\beqar
\Phi^{(q, \mu ,J) }_{mm'~A} (y)&=&
{\sqrt{(2l+1)(2l'+1)}}~\sum_{nn'} {\cal D}^{(l))}_{mn}
(u)~ {\cal D}^{(
l')}_{m'n'} (u') ~\la J A\vert l n; l' n'\ra \nonumber\\
&=& \sqrt{d(q,\mu ,J)}~\D^{(R)}_{(mm') JA} (L^{-1}_y)
\label{11}
\eeqar
where $\la J A\vert l n: l' n'\ra$ denotes the
Clebsch-Gordan coefficient connecting the $SU(2)$ states
$\vert l n\ra$, $ \vert l' n'\ra$ to
$\vert J A\ra$. In the second line of (\ref{11}),
we give the notation in terms of the Wigner function for
$G$, where $R$ designates the representation $(l,l')$
and the right index $JA$ indicates restriction to
the spin-$J$ representation of the subgroup $H=
SU(2)_{diag}$
and the choice of the state $A$ within this
representation. $L_y\in SO(4)$ is a coset
representative of the point $y^\mu \in S^3$.
The orthogonality of the Wigner functions shows that we
have the
normalization condition
\beq
\int_{S^3} d\mu (y)~ \Phi^{*(q,\mu ,J)}_{mm'~A}
\Phi^{(q', \mu' ,J')}_{nn'~B} ~=~\delta^{qq'}\delta^{\mu
\mu'}
\delta^{JJ'} \delta_{mn} \delta_{m'n'}
\delta_{AB}\label{11a}
\eeq
The mode expansion for the fermion field operator is
given by
\beq
\psi_A (y)= \sum_{\mu ,mm'} ~a^{(q, \mu )}_{mm'}
~\Phi^{(q, \mu ,J)}_{mm'~A}
(y)
\label{12}
\eeq
The index $A$ refers to the gauge charge (or isospin)
components
of the field. Here $a^{(q, \mu )}_{mm'}$ are the particle
annihilation
operators, so that the completely filled lowest Landau
level
can be written as
$\prod_{mm'} a^{(0,T)\dagger}_{mm'}\vert 0\ra$, for
example,
for the integral
$T$-case.

We now turn to the relativistic case.
The Dirac Hamiltonian on $S^3$ is given by
\beq
H= \left[ \matrix{ m&{\sigma\cdot K\over 2r}\cr
{\sigma\cdot K\over 2r}&-m\cr}\right]\label{13}
\eeq
where $\sigma_i$ are the Pauli matrices for spin.
As usual, there are positive and negative energy solutions
$E = \pm \omega$, where $\omega$ will be given by
the eigenvalues of $\sqrt{(\sigma\cdot K/2r)^2 +m^2}$.
We will concentrate on the positive energy solutions;
the negative energy solutions will be similar.
For $E =\omega$,
\beq
\psi_A = \left( \matrix{U_A\cr V_A\cr}\right),
\hskip 1in V= {(\sigma\cdot K)\over
2r (\omega +m)}~U
\label{13a}
\eeq
The operator $\sigma\cdot K$ acts
on fields of the form $U_{A\alpha} (g)$ where
$\alpha$ is a spin index.
Using the commutation rules for the $K$'s, we find
$(\sigma\cdot K)^2 = K^2 -\sigma \cdot J$.
We will combine the spin with the isospin to form $J_i$.
In doing so, notice that $\sigma_i$ act on the left
as $(\sigma_i U)_\alpha = (\sigma_i)_{\alpha \beta}
U_\beta$ while
the $T$'s act on the right. So we write $\sigma U =
U \sigma^T = -2~U~ S_i$, identifying the spin
$S_i$ as $-\half \sigma_i^T$ (which does obey the standard
commutation rules).
Thus
\beq
(\sigma \cdot K)^2 = K^2 + 2S \cdot J\label{14}
\eeq
The last term represents the Zeeman effect.
The energy eigenvalues are then given as
\beqar
\omega &=& \left[ {(\sigma \cdot K )^2 \over 4r^2} ~+~m^2
\right]^{1\over 2}
\nonumber\\
&=& \left[ {{(q+J+1)^2 +(J-\mu )^2 -1 +S(S+1) -
T(T+1)}\over 4r^2}
~+~m^2\right]^{1\over 2}\label{15}
\eeqar
The degeneracy is as before $d (q,\mu ,J)=
(q+ \mu +1) (q-\mu +2J +1)$.
With $S=\half$, the allowed values of $J$ for a given
isospin
$T$ are $T\pm \half$.

If $T$ is an integer, the lowest states correspond to
$q=0$, $J=T-\half$, $\mu = T$ and $\mu = T-1$.
$(\sigma\cdot K)^2
= {1\over 4}$ and the degeneracy for each tower
is $d = T (T+1)$. If $T$ is half an odd integer, the
lowest
state corresponds to $q=0$,
$J= T-\half$, $\mu =J= T-\half$ with
$(\sigma\cdot K)^2 =0$ and degeneracy $d = (T+\half )^2$.

In working out the mode expansion for the fermion fields,
we can first split $\psi$ into the $J =T+\half$ and
$J= T-\half$ pieces, for each of which one has the
$\Phi^{(q, \mu ,J)}_{mm'}$
solutions of (\ref{11}). The spinorial solutions are thus
\beq
\Psi^{(q, \mu , T \pm \half ) }_{mm'~A}=
\left( \matrix{ U^{(q, \mu , \pm )}_A\cr
V^{(q, \mu ,\pm )}_A\cr}\right)_{mm'}\label{16}
\eeq
\beqar
U^{(q, \mu , +)}_{A\alpha} &=&
\sum_B \la T A; \half \alpha \vert T+\half ~B\ra
~\Phi^{(q, \mu ,T+\half )}_{mm'~B}
(g, h)\nonumber\\
U^{(q, \mu , -)}_{A\alpha} &=&
\sum_B \la T A; \half \alpha \vert T-\half ~B\ra
~\Phi^{(q, \mu ,T-\half )}_{mm'~B}
(g, h)\label{17}
\eeqar
The mode expansion is thus
\beqar
\psi_A &=& \sum_{q, \mu ,mm'}~
a^{(q,\mu ,T+\half )}_{mm'} \Psi^{(q,\mu ,T+\half
)}_{mm'~A}
+ a^{(q,\mu ,T-\half )}_{mm'} \Psi^{(q,\mu ,T-\half
)}_{mm'~A}\nonumber\\
&&\hskip 1.5in +{\rm negative ~energy ~part}\label{18}
\eeqar
The lowest Landau level may again be written in terms
of the particle creation operators $a^{(q,\mu ,T- \half
)\dagger}_{mm'}$.

\section{The effective action for edge states}

We now turn to quantum Hall droplets and the nature
of the edge excitations.
In making up a quantum Hall droplet,
we will be filling
up all the negative energy states and a large number
of positive
energy states.
The energy of the edge states involves the difference of
eigenvalues
near the boundary of the droplet and there is no
qualitative
difference between the relativistic and nonrelativistic
cases
for this; therefore, for most of the rest of our analysis,
we shall only consider the nonrelativistic case.

On ${\bf CP}^k$ spaces with a $U(1)$ magnetic field the
density of
states scales like the volume of space in the large volume
limit
\cite{KN1}.
So for every unit cell of volume corresponding to the
magnetic
length, one has exactly one state. A droplet of fermions
is then
incompressible and the only low energy excitations are
volume-preserving deformations of the droplet, i.e., edge
excitations.
In the case of $S^3$, the situation is more complicated.
From the energy level formula (\ref{10}) we
see that, in the large $J$-limit, the splitting of energy
levels
is finite if $J\sim r^2$, as $r\rightarrow \infty$.
The same result holds for the relativistic case for states
near $\mu \approx J$; this is easily checked from
(\ref{15}).
(States near $\mu \approx 0$ or $\mu \approx 2 J$ can be
infinitely
separated.)
For the degeneracy of a Landau level, we have
$d\sim r^4$. Thus the number of states per unit volume of
$S^3$, namely $d/V$, diverges as $r$.
This situation is somewhat similar to what happens on
$S^4$ with an instanton field where $d/V \sim r^2$
\cite{HZ}.
A nice interpretation of the latter result is to consider
a $U(1)$ magnetic field on ${\bf CP}^3$ which is an
$S^2$-bundle
over $S^4$. The $U(1)$ magnetic field
is the instanton from the $S^4$ point of view. Since
$d/vol({\bf
CP}^3)\sim$ constant, one can consider droplets of uniform
density
on ${\bf CP}^3$, leading to incompressible states and the
usual edge states.
This was suggested in \cite{KN1}, and analyzed and
explained
in detail in \cite{ners, berneveg}. The effective
Lagrangian for edge states
on $S^4$ has also been obtained via this construction
\cite{KN2}.

One can attempt a similar interpretation for $S^3$ by
starting with
$S^2\times S^2$. To see this connection, consider the
states
$\Phi^{(q,\mu ,J)}_{mm'~A}(y)$ given in (\ref{11}), say,
for
the lowest Landau level, so that
$q=0$ and $\mu =J=T$, taking integer
$T$ as an example. The magnetic translation operators form
$SU(2)\times SU(2)$, corresponding to the $SU(2)$
operators
$L_a,~L'_a$ acting as linear transformations on the
indices
$m,~m'$ respectively.
We will take the Hamiltonian to be now given by
(\ref{6}) with a potential $V$ added; the potential
$V$ is a function of $L_a$ and $L'_a$ which breaks the
magnetic
translation symmetry and localizes the fermions
in some region of the Hilbert space. The localized
fermions form the droplet. The occupied states are
specified in
terms of a density matrix ${\hat \rho}_0$.
One can then work out an effective action for the
edge excitations following the procedure
outlined in \cite{sakita, KN2}. The dynamical evolution of
${\hat \rho}_0$, which includes all the edge excitations,
is described by a unitary matrix ${\hat U}$,
${\hat \rho}_0 \rightarrow {\hat U} {\hat \rho}_0
{\hat U}^\dagger$.
The action for ${\hat U}$ is given by
\beq
S = i \Tr \left({\hat \rho}_0 {\hat U}^\dagger \del_t{\hat
U}
\right) ~-~ \Tr ({\hat \rho}_0 {\hat U}^\dagger {\hat V}
{\hat U})
\label{19}
\eeq
The dynamical degrees of freedom for excitations around
the chosen
${\hat \rho}_0$ are in ${\hat U}$. An effective action for
excitations of the droplet can be obtained by simplifying
the action (\ref{19}) in the large $J$ limit.

The basic strategy in simplifying this
action is to approximate the operators
${\hat U}, ~{\hat V}$, etc. by $c$-number
functions in the limit of a large number of states.
In doing so, we will also
encounter commutators which can be replaced by
appropriate Poisson brackets. In the case of
the ${\bf CP}^k$ spaces considered in \cite{KN2},
the Poisson brackets are defined by the
K\"ahler structure on ${\bf CP}^k$.
The key to extracting the Poisson limit of the commutators
is the star
product, or at least the first two terms of such a
product,
and so we will start with this question.
As mentioned before, the wavefunctions undergo gauge
transformations
corresponding to the chosen representation of $H$. Our
approximations
should respect this. Our procedure
will be to start with
the Wigner functions and define symbols or classical
functions associated
with any operator (or matrix) on the Hilbert space of LLL
states.
The symbols
will turn out to be $(2J+1)\times (2J+1)$-matrices,
rather than functions, so as to keep gauge covariance
of the products.
We can then proceed to the large $J$ limit,
which
will give us the required terms in the effective action,
but will still involve traces of matrix products, where
the matrices are of dimension $2J+1$, with
$J\rightarrow \infty$. As a second step,
one can approximate
these traces by integrations over classical functions as
well.
We are then naturally led to $S^2 \times S^2$.
This second step of approximating the trace over the
representation of the gauge group
will break gauge invariance in general, except for
certain choices of the density matrix, which are the
analogues
of the density matrix chosen for the ${\bf CP}^3$ to $S^4$
reduction.

Turning to the details, notice that for
the lowest Landau level, the dimension of the Hilbert
space is
$d= n^2$, $n = 2l+1 =J+1$. The observables are thus in the
Lie algebra
of $U(n^2)$. A basis for this can be taken as the $n^4$
matrices given by $1, L_a, L'_a$ and all independent
products thereof. Consider a typical matrix ${\hat A}$
with matrix
elements $A_{\alpha \beta}= A_{mm' ,nn'}$.
We have an $SU(2)$ background field on $S^3$ and the
wavefunctions
$\Phi_{mm'A}$ are sections of an $SU(2)$-bundle on
$S^3$, as is evident from their gauge transformation
property.
Therefore the symbol corresponding to ${\hat A}$ is a
matrix-valued function on $S^3$ and we define it as
\beq
A_{AB}(y)=
\sum \D^{(R)}_{JA~\alpha }(L_y)~A_{\alpha \beta}
~\D^{(R)}_{\beta ~JB}(L^{-1}_y)
\label{20}
\eeq
where in the second line we have given the compact
notation in terms of
a single $G$-representation.
$R= (l,l)$ and $\alpha ,\beta$, etc., are composite
indices
and the right indices $JA$ and $JB$ indicate restriction
to
the $H$-representation of spin $J$ and the states $A,B$
within
that representation.
We can write this in a more compact way as
\beq
A_{AB}(g)=
\sum_{\alpha \beta} \la JA\vert L_y\vert \alpha\ra
~A_{\alpha \beta}
~\la \beta\vert L^{-1}_y\vert JB\ra
\label{21}
\eeq
$H$-transformations act on $L_y$ as $L_y\rightarrow h L_y$,
$h\in H$. Under such a transformation, the symbol
for ${\hat A}$ transforms as
\beq
A'_{AB}(y)= h_{AC} ~A_{CD}(y)~ h_{DB}^\dagger
= (h A h^\dagger)_{AB} \label{21a}
\eeq

The trace of the operator ${\hat A}$ is given by
\beqar
\Tr {\hat A} &=& \sum_{\alpha} A_{\alpha \alpha}
= \sum_{\alpha\beta} A_{\alpha \beta}~
{(J+1)^2 \over 2J+1} \int_{S^3} d\mu (y)~
\D^{*(R)}_{\alpha ~JA} (L^{-1}_y)
\D^{(R)}_{\beta ~JA} (L^{-1}_y) \nonumber\\
&=& {(J+1)^2 \over 2J+1}\int_{S^3} d\mu (y)~ \tr A(y)
\label{22}
\eeqar
On the right hand side, we have the trace over the
$H$-representation
and the integral over $L_y$. The Haar measure $d\mu$
is normalized to
unity.

Consider now the product of two matrices ${\hat A}$ and
${\hat B}$ and
symbol corresponding to it. We can write
\beqar
({\hat A}{\hat B})_{AB}(y)
&=& \la A \vert L_y\vert \alpha\ra A_{\alpha\beta}B_{\beta
\gamma}
\la \gamma \vert L^{-1}_y\vert B\ra\nonumber\\
&=& \la A \vert L_y\vert \alpha\ra A_{\alpha\beta}
\la \beta \vert L^{-1}_y L_y\vert \beta'\ra B_{\beta'
\gamma}
\la \gamma \vert L^{-1}_y\vert B\ra\nonumber\\
&=& \sum_{j r} \la A \vert L_y\vert \alpha\ra
A_{\alpha\beta}
\la \beta\vert L^{-1}_y\vert j r\ra~ \la j r\vert L_y
\vert
\beta' \ra B_{\beta' \gamma}
\la \gamma \vert L^{-1}_y\vert B\ra\label{23}
\eeqar
There is summation over indices like $\alpha,~ \beta$,
$\beta' ,~\gamma$ in
these formulae.
We have explicitly indicated the summation over
intermediate states
in the last line of this equation. This summation runs
over
all representations of the subgroup $H$, i.e., over all
$j= J, J-1, ...,0$, and over all states
$r$ within each such representation.
The term corresponding to $j =J =2l$ will give the
product of the symbols (classical functions) for ${\hat
A}$ and
${\hat B}$ considered as $(2J+1)$-dimensional matrices.
This is the first term of the star product.
The remaining terms can be simplified by noting that
there is an operator $\Lambda_a$ which
can map from the $j =J$ representation
to the $j =J-1$ representation, i.e., a lowering operator
for the $H$-Casimir ${\hat J}^2$.
$\Lambda_a$ is explicitly given by
\beq
\Lambda_a = K_a ~\left(1 - \sqrt{4 {\hat J}^2 +1}\right)
+2i \ep_{abc} K_b J_c
\label{24}
\eeq
(Whenever it is necessary to specify that
$J$ is an operator rather than the $c$-number $J=2l$,
we use ${\hat J}$; of course,
$J_a$ with the subscript is always an
operator.) $\Lambda_a$
obeys the commutation rule
\beq
[J^2 , \Lambda_a ] = \Lambda_a ~\left( 1 - \sqrt{4{\hat J}^2
+1}\right) -4J_a K\cdot {\hat J}
\label{25}
\eeq
The second term on the right hand side of (\ref{25})
 gives zero on the LLL states while the first term shows
that
  $\Lambda_a \vert j, A\ra$ is some state
with spin $j-1$. ($\Lambda^\dagger_a$ can be used to
raise the $j$-value.)

Starting with the $SU(2)\times SU(2)$ representation
$(l,l)$, the $H$-representations have spin values
$j= 0,1,\cdots, 2l$, each occuring once.
Thus the operator $\sum_{aA} \Lambda_a \vert j A\ra
\la j A\vert\Lambda^\dagger_a$ is zero for all states
of the LLL Hilbert space
except on the $j-1$ subspace. Since $\la j-1 ~M \vert
\Lambda_a
\vert j' A\ra =0$ for all
$j'$ except for $j'=j$, we have
\beq
\sum_{aA} \la j-1 ~M\vert \Lambda_a \vert jA\ra \la
jA\vert
\Lambda^\dagger_a \vert j-1 ~N\ra
=\la j-1 ~M\vert \Lambda_a \Lambda^\dagger_a \vert j-1
~N\ra
\label{26}
\eeq
Further, since $J_i$ commutes with $\Lambda_a
\Lambda^\dagger_a$,
the latter operator must be proportional to the identity
on these states. So we can write
\beqar
\la j-1 ~M \vert \Lambda_a \Lambda^\dagger_a \vert
j-1 ~N\ra &=& \delta_{MN} \la \Lambda^2\ra\nonumber\\
\la \Lambda^2\ra &\equiv& \la j-1~ j-1\vert \Lambda_a
\Lambda^\dagger_a
\vert j-1~ j-1\ra
\label{27}
\eeqar
For the highest $H$-representation with
$j= J$ we find
\beq
\la \Lambda ^2\ra = 4J {(2J+1)^3 \over 2J-1}
\equiv C(J)\label{28}
\eeq
Putting all this together, we find that the completeness
relation
may be written as
\beq
\sum_{A} \vert JA\ra \la JA\vert
~+~\Lambda_a \vert JA\ra~ {1\over C(J)}~\la JA\vert
\Lambda^\dagger_a
+\cdots ={\bf 1}
\label{29}
\eeq
We can use this to write out the various terms in the sum
in (\ref{23}).
Finally, notice that $\Lambda_a$ can be represented as
differential
operators on the group elements which are the argument
of
the $\D$-functions in (\ref{23}).We will use ${\hat
\Lambda}_a$
to denote $\Lambda_a$ as a differential operator.
Expanding the sum in (\ref{23}) we can then write the
product as
\beqar
({\hat A}{\hat B})_{AB}(y) &=& A_{AC}(g) B_{CB}(y)
~+~{1\over C(J)}
{\hat\Lambda}_a A_{AC}(y)~ {\hat \Lambda}^\dagger_a
B_{CB}(y)
~+\cdots \nonumber\\
&=&\bigl( A(y)* B(y)\bigr)_{AB}
\label{30}
\eeqar
This gives a version of the star product we need.
(The higher terms in the sum can also be written
using multiple applications of $\Lambda_a$ and
$\Lambda^\dagger_a$.)
For the commutator of two operators, we get
\beqar
([{\hat A}, {\hat B}])_{AB}&=& [A,B]_{AB}+
{1\over C(J)}\biggl(
{\hat\Lambda}_a A_{AC}(y) ~{\hat \Lambda}^\dagger_a
B_{CB}(y)
~-~{\hat\Lambda}_a B_{AC}(y)~ {\hat \Lambda}^\dagger_a
A_{CB}(y)
\biggr)+\cdots\nonumber\\
&=& [A,B]_{AB}+{i\over J} \{ A, B\}_{AB}~+\cdots\nonumber\\
\{A, B\}_{AB}&\equiv&-i {J\over C(J)}
\biggl(
{\hat\Lambda}_a A_{AC}(y)~ {\hat \Lambda}^\dagger_a
B_{CB}(y)
~-~{\hat\Lambda}_a B_{AC}(y)~ {\hat \Lambda}^\dagger_a
A_{CB}(y)
\biggr)
\label{31}
\eeqar
The action of the operator $\hat\Lambda_a$ can be represented as
a differential operator on the symbol.
It is given by
\beqar
{\hat \Lambda}_a A_{AC}(y) &=&
\langle A\vert L_y \hat A L^{-1}_y \Lambda_a\vert
C\rangle\nonumber\\
&=&  -2i J \left[ \left( \delta_{ab}\delta_{CD}-{i\over J}
\varepsilon_{abc}(J_c)_{DC}\right)
 ~{\hat \nabla}_b A_{AD} ~+{\cal O}(1/J)\right]
\label{35a}
\eeqar
where the covariant derivative is defined by
\beq
-2 i{\hat \nabla}_a A_{AD} = \la A \vert~ [K_a ,L_y {\hat A} L^{-1}_y]
~\vert D\ra
\label{35b}
\eeq
This is shown in the appendix. ${\hat \nabla}= r \nabla$ involves only
angular derivatives.

Equation(\ref{31}) does not strictly define a Poisson
bracket because of the order of
the matrix multiplication, but we will use the same
notation.
In this equation, a part of the summation over states
has been reduced in a form suitable for large
$J$ expansion; the coefficient $C(J)$ will give
suppression by powers of $J$.
However, the summation over the gauge indices remain
and they also range over an infinity of values as
$J\rightarrow
\infty$.
Notice that since the $R_a, R'_a$
are covariant derivatives, various terms in this expansion
are gauge covariant. If we truncate the summation over
the gauge indices, we will lose this invariance in
general.

Now we can go back to the action (\ref{19}) and write
${\hat U}= \exp (i{\hat\Phi} )$ and expand in powers of
${\hat\Phi}$. We then find terms like $[{\hat\Phi} , {\hat
\rho}_0]$;
the commutator can be replaced, in the large $J$-limit by
the
bracket $\{\Phi , \rho_0 \}$. We will get an effective
matrix action,
with the summation over the gauge indices.

It is useful at this stage to consider the
nature of the density $\rho_0$ which is a matrix
$(\rho_0)_{AB}$. If all the states are filled, then
${\hat \rho}_0 ={\bf 1}$ and $(\rho_0)_{AB}=\delta_{AB}$.
There can also be other cases where $\rho_0$ is invariant
under $H$-transformations. This means that the overall
charge of the droplet is zero.
If this is not the case, then there are collective
excitations
which are charge rotations which upon quantization lead to
a complete charge multiplet of states.
These charge rotations are generated by spatially
constant $H$-transformations which can depend on time.
They are not gauge transformations.
$H$-transformations
which may depend on the coordinates but which are
time-independent
correspond to gauge transformations.
The procedure for simplifying the action which we have
outlined above
applies to a general choice of ${\hat \rho}_0$
which need not be $H$-invariant.
For $H$-invariant choices of ${\hat \rho}_0$ we can do
a further simplification, converting the remaining sum
over the gauge indices to an integration as well.
We will now show how this case works out.

The states which are
being summed over in the remaining matrix products 
in (\ref{30}, \ref{31}) have
$j =J$; among these there is the highest weight state
$\vert JJ\ra = \vert ll,ll\ra$.
We can expand the matrix products around this highest
weight
state using
\beqar
\sum_A \vert JA\ra \la JA\vert &=&
\sum_s {(2J-s)! \over (2J)! s!}~J_-^s \vert JJ\ra \la
JJ\vert J_+^s
\nonumber\\
&=& \vert JJ\ra \la JJ\vert + {1\over 2J} J_-\vert JJ\ra
\la JJ\vert
J_+ +\cdots \label{32}
\eeqar
This shows that it is convenient to use
a different definition for the symbol of an operator;
we define the
new symbol for an operator as
the scalar quantity
\beqar
({\hat A})(y)&=& A(y) = A_{JJ}(y)\nonumber\\
&=& \sum_{\alpha \beta} \la JJ\vert
L_y \vert \alpha \ra A_{\alpha \beta} \la \beta \vert
L_y^{-1} \vert JJ\ra
\label{33}
\eeqar
Note that $A(y)$ is invariant under $U(1)_R\times
U(1)_{R'}$
(generated by $R_3$ and $R'_3$)
so that this is indeed the symbol we would
define for
$S^2\times S^2$.
Under an $H$-transformation,
 this does not transform covariantly
but mixes with all of $A_{AB}(y)$. The symbols
(\ref{33}) are invariant under the $U(1)$ subgroup of $H$
defined
by $J_3$. Eventhough $A(y)$ is not $H$-covariant,
we can write the trace of an operator as
\beqar
\Tr {\hat A}&=& (J+1)^2 A_{\alpha \beta} \int d\mu (y)
\la JJ\vert L_y\vert \alpha\ra \la \beta \vert
L_y^{-1} \vert JJ\ra \nonumber\\
&=& (J+1)^2 \int d\mu (y) ~A(y)\label{35}
\eeqar
The integration in (\ref{35}) is over $S^2\times S^2$,
not on $S^3$.

For the product of two operators, we find the star product
\beqar
({\hat A}{\hat B})(y)&=& A(y)* B(y) \nonumber\\
&=& \sum_{ss'} (-1)^{s+s'}\left[ {(J-s)! ~(J-s')! \over J!
s! J! s'!}
\right] R_-^s {R'}_-^{s'} A(y)~ R_+^s {R'}_+^{s'}
B(y)\nonumber\\
&=& A(y) B(y) - {1\over J} \biggl( R_-A R_+B ~+~ R'_-A
R'_+ B\biggr)+
\cdots\nonumber\\
&=& A(y) B(y) -{1\over 2J} \left( J_-A J_+B ~+~ K_-A K_+B
\right)
+\cdots \label{36}
\eeqar
(The explicit formula for $R_{\pm}$ is $R_{\pm}= 
e_{\pm}^\mu\partial_\mu$,
where $e^\mu _{\pm} = e_1 ^\mu +ie_2 ^\mu$, with $e_a^\mu$ being an
orthonormal frame on $S^2$. Similar formulae are valid for$R' _{\pm}$ on
$S'^2 $.)

For the commutator, we get
\beqar
\left( [{\hat A}, {\hat B}]\right) (y) &=&
{i\over J} \biggl[ \{ A, B\}_J ~+~ \{ A, B\}_K \biggr]
\equiv{i\over J} \{ \!\!\{ A, B\}\!\! \}\nonumber\\
\{ A, B\}_J &=& {i\over 2} (J_-A J_+B - J_-B J_+A
)\nonumber\\
\{ A, B\}_K &=& {i\over 2} (K_-A K_+B - K_-B K_+A
)\label{37}
\eeqar

The effective action can now be simplified using the large
$J$
relations given here. The result is
\beq
S \approx - {(J+1)^2 \over 2J} \int d\mu ~
\biggl[ \{ \!\! \{ \rho_0 , \Phi \}\!\!\} \del_t \Phi
~+~ \{ \!\! \{ \rho_0 ,\Phi \}\!\! \} \{\!\! \{ V, \Phi \}\!\! \}
\biggr]\label{38}
\eeq
$J$'s are the generators of
the $H$-transformations and
if $V$ and $\rho_0$ are $H$-invariant,
the $J$-brackets vanish for these quantities
and the action simplifies to
\beq
S \approx - {(J+1)^2 \over 2J} \int d\mu ~
\biggl[ \{ \rho_0 , \Phi \}_K \del_t \Phi
~+~ \{ \rho_0 ,\Phi \}_K \{ V, \Phi \}_K
\biggr]\label{39}
\eeq

This result involves derivatives $K_\pm$, which may seem to be
a specific choice of directions. There is nothing special
about
this choice. We could equally well have expanded around
the highest eigenvalue of
$J\cdot {\hat e}_3$, for some unit vector
${\hat e}_3$ rather than $J_3$. $K_\pm$ are then replaced
by $({\hat e}_1 \pm i {\hat e}_2 )\cdot K$, where
${\hat e}_i$ form a triad of orthonormal vectors.
This degree of freedom is already contained in
the variables we are using. It is easy to see that
any rotation of the frame ${\hat e}_i$ can be absorbed
into $L_y$ as a transformation $L_y\rightarrow h L_y $, $h\in H$.
Thus the action (\ref{39}) will lead to rotationally
invariant results.

The action (\ref{39}) is the precise equivalent for $S^3$
of the situation for $S^4$ obtained from ${\bf CP}^3$
via the choice of a local
complex structure. To complete this analogy, we must now
consider
the question of what kind of potential will lead to
a density that is invariant under $H$-transformations.
The potential $V$ must be a function of the magnetic
translation operators $L_a, ~L'_a$. We choose it to be
of the form
\beq
V= \lambda \left[ J(J+1) - (L+L')^2\right] \label{40}
\eeq
The effect of this potential is to fill states
as multiplets of the algebra of $J^L_a=
L_a +L'_a$,
starting from the highest value $J(J+1)$.
Assume that a certain number of
multiplets, say $J$, $J-1, \cdots, J-M$,
have been filled.
The symbol for density associated
with this choice is
\beq
(\rho)_{AB} = \sum_{j=J}^{J-M}\sum_r
\la JA \vert L_y \vert j r\ra \la jr\vert L^{-1}_y\vert JB\ra
\label{41}
\eeq
Introduce local coordinates by writing $L_y = S V$,
where $V$ is an element of the $H$-subgroup defined
by $J^L_a$. Since we have complete $H$-multiplets
for the states $\vert jr\ra$ in the sum in
(\ref{41}), $V$'s cancel out. By expanding
$S, S^{-1}$ in a series of the operators 
$\Lambda_a,
\Lambda^\dagger_a$ which lower and
raise
the $j$-value, 
and using an argument similar to how we
arrived at (\ref{27}), we can see that $(\rho_0)_{AB}$ is
indeed proportional to $\delta_{AB}$.

\section{$S^2\times S^2$ and fuzzy $S^3/{\bf Z_2}$}

The analysis we have done for extracting an effective
action for edge
states led to $S^2\times S^2$.
The latter plays a role analogous to what ${\bf CP}^3$
does for
$S^4$.
All the operators of
interest transformed as integral spin representations of
$H$, so
$S^3/{\bf Z}_2$ is adequate for most of what we have done.
The space
$S^3/{\bf Z}_2$ can be embedded in $S^2\times S^2$. The
latter space
can be described by $n^2 =1$, $m^2=1$,
$n =( x_1, x_2, x_3)$, $m =( y_1, y_2, y_3)$.
The space $S^3/{\bf Z}_2$ is now obtained by imposing
the further condition $n\cdot m = x_1y_1 +x_2 y_2 +x_3y_3
=0$. It is clear that any solution to these
equations gives an $SO(3)$ matrix $R_{ab}=
(\ep_{abc}m_bn_c,
m_a, n_a)$. Conversely, given any element $R_{ab}\in
SO(3)$,
we can identify $n_a = R_{a3}$, $m_a =R_{a2}$. There are
other ways
to identify $(n,m)$ but these are equivalent to
choosing different sets of values
for the $SO(3)$ parameters; this statement can
be easily checked using the Euler angle parametrization.
What we have described is essentially the angle-axis
parametrization of rotations \cite{tung}.
Since the
space $S^2\times S^2$ has K\"ahler structure,
it is the simplest enlargement of space we can use
to
define coherent states and large $J$ limits.
Further since $vol(S^2\times S^2)\sim r^4$, $d/V$
goes to a constant as $r\rightarrow \infty$,
so
we can get incompressible Hall droplets just as in the
$S^4$-${\bf CP}^3$ case.
This is basically what we have utilized.

There is an interesting connection between
lowest Landau level states
and the fuzzy version of the space on which the Landau
problem is defined. For the Landau problem
on ${\bf CP}^k$, one can consider the set of all
hermitian operators on the LLL Hilbert space.
This will correspond to the generators of
$U(d)$ where $d$ is the dimension of the
LLL Hilbert space. These operators, in the large
$d$ limit, are in one-to-one correspondence with
the basis of functions on ${\bf CP}^k$.
Thus, at finite $d$, operators on
the LLL Hilbert space provide a fuzzy version
of ${\bf CP}^k$.
Since we have defined the Landau problem on $S^3$
we can now ask the question whether
a similar relation is obtained here.
It will turn out that there is
some relation with fuzzy $S^3/{\bf Z}_2$,
not quite so simple as in the even dimensional cases.
To see this,
we need to first consider a fuzzy version
of $S^3/{\bf Z}_2$.

Consider $SU(2)\times SU(2)$, with generators
$L_a , L'_a$ and take a particular representation
where $l=l'$, so that we can think of $L, L'$ as
$(n\times n)$-matrices, $n=2l+1$.
Since the quadratic Casimirs $L^2 =L'^2 = l(l+1)$,
this gives
the standard realization of fuzzy $S^2 \times S^2$
\cite{madore}.
As $l$ becomes large, we can use the standard coherent
state
representation of $SU(2)/U(1)$ to show that
\beq
L_a \approx l ~2 \Tr (g^\dagger t_a g t_3) ,
\hskip 1in
L'_a \approx l ~ 2 \Tr ( g'^\dagger t_a g' t_3)
\label{42}
\eeq
where $g, g'$ are $(2\times 2)$-matrices
parametrizing the two $SU(2)$'s and
$t_a =\half \sigma_a$, $\sigma_a$ being the Pauli
matrices.
(All functions of $L, L'$ are similarly
approximated.)
To get to a smaller space, clearly we need to
put an additional restriction which we will take
as the following.
An operator is considered admissible or physical if
it commutes with $L\cdot L'$, or equivalently
commutes with $(L-L')^2$ or $(L+L')^2$, i.e,
\beq
[ {\cal O}, (L-L')^2 ]=0
\label{43}
\eeq
It is easily seen that
the product of any two operators which obey this
condition will also obey the same condition, so this leads
to a closed algebra.
A basis for the vector space on which $L, L'$ act
is given by $\vert l m l m'\ra$ in the standard angular
momentum notation. We rearrange these into multiplets
of $J^L_a = L_a +L'_a$. For all
state within each irreducible representation
of the $J^L$-subalgebra labelled by $j$, $(L-L')^2$
has the same eigenvalue $4 l(l+1) - j (j+1)$.
Operators which commute with it are thus block diagonal,
consisting of all unitary transformations on
each $(2j+1)$-dimensional subspace.
There are $(2j+1)^2$ independent transformations
for each $j$-value putting them in one-to-one
correspondence
with the basis functions $\D^{j}_{ab}(U)$ for an
$S^3$ described by the $SU(2)$ element $U$.
By construction, we get only integral values of
$j$, even if $l$ can be half-odd-integral,
so we certainly cannot
get $S^3$ in the large $l$ limit, only
$S^3/{\bf Z}_2$.

We can go further and ask how the condition
(\ref{43}) can be implemented in the large
$l$ limit. This can be done by fixing the value of
$L\cdot L'$ to be any constant. Using (\ref{42})
we find that this leads to
\beq
L\cdot L' \sim 2 \Tr ( g'^\dagger g ~t_3 g^\dagger
g' t_3 )
\sim {\rm constant}
\label{44}
\eeq
This means that
\beq
g'^\dagger g = M \exp(it_3 \gamma )
\label{45}
\eeq
where $M$ is a constant $SU(2)$ matrix.
$\gamma$ can be absorbed into $g$.
Since $L\cdot L'\sim
2 \Tr ( M t_3 M^\dagger t_3)$, we can take
$M = \exp (it_2 \beta_0)$ using the Euler
angle parametrization. We then find
\beqar
L_a &\sim& 2 \Tr (g^\dagger t_a g t_3 )\nonumber\\
L'_a &\sim& \cos\beta_0 ~2 \Tr (g^\dagger t_a g t_3)
+\sin\beta_0~ 2 \Tr (g^\dagger t_a gt_1)\label{46}
\eeqar
Thus all functions of these can be built up from the
$SO(3)$ elements $R_{ab}= 2 \Tr (g^\dagger t_a gt_b)$.
(Actually we need $b=1,3$, but $b=2$ is automatically
given by the cross product.)
Thus, in the large $l$ limit, the operators obeying
the further condition (\ref{43}),
tend to the expected mode functions for the group
manifold of $SO(3)$ which is $S^3 /{\bf Z}_2$.
We have thus obtained a fuzzy version of
$S^3/{\bf Z}_2$ or ${\bf RP}^3$.
The condition we have imposed, namely (\ref{43}), is
also very natural, once we realize that $(L-L')^2$
is the matrix analogue of the Laplacian, and
mode functions can be obtained as eigenfunctions
of the Laplacian.

Noncommutative, but not fuzzy,
spheres and real projective spaces have been obtained
before \cite{landi};
for considerations related to fuzzy spheres,
see \cite{ram}.

It is clear from our discussion that the LLL
states on $S^3$ do not lead to just fuzzy $S^3/{\bf Z}_2$.
We do get the set of operators needed for fuzzy $S^3/{\bf
Z}_2$,
but there are more. The operators which do not
obey (\ref{43}) are physical operators for the Landau
problem. They are needed
if we attempt to describe noncommutative
algebra of functions on fuzzy $S^3 /{\bf Z}_2$
via star products.

\section{Discussion}

We have carried out the analysis of the Landau problem
on $S^3$ taking a constant background field proportional
to
the spin connection on $S^3$. One can tune the coupling
to the gauge field by changing the dimension of the
representation
of the charge algebra to which the fermions belong, in
a way precisely analogous to what was done in \cite{HZ}.
We have also obtained the effective action for the edge
excitations of a quantum Hall droplet in the limit of
large fermion representations.
The appropriate space for these considerations is
naturally
enlarged to $S^2 \times S^2$; this is again the
precise analogue of obtaining edge dynamics on
$S^4$ by starting with ${\bf CP}^3$.

We have also given a method of representing the
algebra of functions on fuzzy $S^3/{\bf Z}_2$
and related this to the Landau problem.

We close with some comments on the background field.
Constant gauge fields in a nonabelian theory can have
unstable (tachyonic) fluctuations. This can be seen
in the present case by writing
$A_i^a = a_i^a+V_i^a$,
where $a_i^a$ denotes the background potential and $V_i^a$
are
fluctuations. The quadratic fluctuation term of the
Yang-Mills
action, apart from the time-derivative terms, is
\beq
S^{(2)} = \int \left[ {1\over 4} (\nabla_i V_j -\nabla_j
V_i)^2
+2 f^{abc} F_{ij}^a V_i^b V_j^c\right]
\label{d1}
\eeq
where $F_{ij}^a = -\ep_{ija}$ is the background field. In
the
background gauge $\nabla_i V_i =0$, this may be simplified
as
\beqar
S^{(2)} &=& {1\over 2} \int ~V_i^a (E_G^2)_{ij}^{ab}
~V_j^b
\nonumber\\
E_G^2 &=& K^2 +S^2 + 2 S\cdot T
\label{d2}
\eeqar
Here $S_{ij}^a= -i\ep_{aij}$ is the spin matrix for
vectors.

For the vector field, $S_k,~T_k$ belong to the spin-$1$
representation
of $SU(2)$ and so $J=0, 1, 2$ giving $9$ possible towers
of states. In
this case
\beq
E_G^2 = (q + J + 1)^2 ~+~ (J-\mu )^2 ~-~3
\label{d3}
\eeq
The lowest state ($q = 0$) of the $J=0,~\mu =0$ tower is a
tachyon with
$E_G^2=-2$ in units of the inverse radius of $S^3$.

We can now ask the question: does this vitiate the very
premise of our
analysis, of starting with a constant background field?
We expect the answer is no, because we have a large number
of
fermions coupling to this. The quantum corrections to the
gauge boson
mass from fermions is determined by 
the index
$A_R$ given by $\Tr (t_a t_b)_R = A_R \delta_{ab}$.
(The fermion one-loop contribution is proportional to
$A_R$.)
When the fermion representation becomes very large, this
can
easily overcome any instability for the gauge fields,
although a one-loop calculation is not adequate to prove this
point.
So we expect that
the potential instability should not be a problem.

\section*{Appendix}

In this appendix, we show how the action of
${\hat \Lambda}_a$ can be represented asa differential
operator on the symbols.
The key observation
is that, for the first nontrivial term in (\ref{31}),
we only need the
leading terms in
$1/J$, since we already
have the
$C(J)$ factor.
$\Lambda_a \vert \psi\ra$ for $\vert \psi\ra =
L_y A L_y^{-1}\vert C\ra$ can have a maximal $j$-value of
$J-1$. The state $\la A\vert$ has a $j$-value of $J$.
As a result, we have
\beq
\la A\vert \Lambda_a \vert \psi \ra =0
\label{a1}
\eeq
We can thus write
\beq
\la A \vert L_y A L_y^{-1} \Lambda_a \vert C\ra
= \la A \vert [L_y A L_y^{-1}, \Lambda_a] \vert C\ra
\label{a2}
\eeq
We must now calculate the commutator. Write
$\Lambda_a = K_a f ({\hat J}) + 2i \ep_{abc}K_b J_c$
where $f ({\hat J}) = 1 - \sqrt{ 4 {\hat J}^2 +1}$.
The commutator involves the four terms
\beqar
[L_y A L_y^{-1}, \Lambda_a] &=& - K_a [ f ({\hat J}) , L_y A L_y^{-1}]
- [K_a , L_y A L_y^{-1}] f ({\hat J}) \nonumber\\
&&\hskip .3in+ 2i \ep_{abc}
[L_y A L_y^{-1}, K_b] J_c + 2i \ep_{abc}
K_b[L_y A L_y^{-1}, J_c]
\label{a3}
\eeqar
The first term can be rewritten as follows.
\beqar
{\rm Term ~1}&=& \la A \vert K_a f ({\hat J}) L_y A L_y^{-1} \vert C\ra
- \la A \vert K_a L_y A L_y^{-1} f ({\hat J})\vert C\ra\nonumber\\
&=& \la A \vert [K_a ,f ({\hat J})]~ L_y A L_y^{-1} \vert C\ra
+\la A \vert  f ({\hat J}) K_a  L_y A L_y^{-1} \vert C\ra\nonumber\\
&&\hskip .3in - \la A \vert K_a L_y A L_y^{-1} f ({\hat J})\vert C\ra
\nonumber\\
&=& \la A \vert [K_a ,f ({\hat J})]~ L_y A L_y^{-1} \vert C\ra
\label{a4}
\eeqar
The last line follows from the fact that the two end states are
eigenstates of ${\hat J}^2$ with the same eigenvalue.
The commutator of $K_a$ with $J_c$ goes like $K$; it lowers
the number of $J$'s
by one, so this term, in the large $J$ limit, is of order $K$.
(Notice that $f({\hat J})$ is of order $J$.)
The second term can be written as
\beqar
{\rm Term~2} &=& \la A \vert  [K_a , L_y A L_y^{-1}] f ({\hat J})
\vert C\ra = -2J \la A \vert  [K_a , L_y A L_y^{-1}]
\vert C\ra\nonumber\\
&\equiv& (-2J) {\hat K}_a A_{AC}(L_y)\label{a5}
\eeqar
where ${\hat K}_a$ is the differential operator.
Notice that this term is order $J$ higher than Term 1.
The third term can be written as
\beqar
{\rm Term~3} &=& 2i \ep_{abc} \la A\vert
[L_y A L_y^{-1}, K_b] J_c \vert C\ra\nonumber\\
&=& - 2i \ep_{abc} {\hat K}_b A_{AD} (L_y) (J_c)_{DC}
\label{a6}
\eeqar
This term involves the matrix elements of $J_c$ and so must be
considered
of order $J$.
The last term can be written as
\beq
{\rm Term~4} = 2i \ep_{abc} \la A\vert K_b [J_c , L_y A L_y^{-1}] \vert
C\ra
\label{a7}
\eeq
This term is of order $K$, because the commutator of
$J$ with any matrix lowers the power of $J$ by one.
(The best way to see this is to consider $L_y A L_y^{-1}$ to be expanded
in powers of $J_a$ and $K_a$. Every commutator with $J_c$ lowers the
power of $J$.)

Thus, of the four terms, two are subdominant. The leading terms can be
gathered as
\beq
\la A \vert L_y A L_y^{-1} \Lambda_a \vert C\ra =
2J ~{\hat K}_a A_{AC}(L_y) - 2i \ep_{abc}~ {\hat K}_b A_{AD}(L_y)
~(J_c)_{DC} ~+{\cal O}(1)
\label{a8}
\eeq
This is the quoted result. It may be worth recalling at this stage that
the
covariant derivatives are defined by
\beq
\left(\nabla_a A(y)\right)_{AB} =
e^\mu_a\left(\partial_\mu A_{AB}-i\left[ \omega_\mu,
A(y)\right]_{AB}\right)
\label{a9}
\eeq
where $e^\mu_a(y)$ are the components of an orthonormal
frame on $S^3$ with $\omega_\mu=\omega^a_\mu J_a$
the corresponding spin connections, which take their
values on the algebra of
$SO(3)_J$.  Clearly this  derivative is covariant with
respect to $SO(3)$ gauge
transformation (\ref{21a}).

\vskip .2in
\leftline{\bf Acknowledgments}

We thank G. Landi for a very useful
conversation.
VPN thanks the Abdus Salam International Centre for
Theoretical
Physics for hospitality during the course of this work. 
VPN was supported in
part by the National Science Foundation
under grant number PHY-0244873 and
by a PSC-CUNY grant.

\end{document}